\documentclass[mathpazo]{cicp}

\usepackage{graphics,graphicx,dcolumn,bm,fleqn,epic,eepic,float,epsfig}
\usepackage{amssymb,amsmath,multirow,rotate,color,float}
\usepackage{epstopdf}
\usepackage{times}
\usepackage{color}
\usepackage{soul}                    % package needed for overstriking
\definecolor{red}{rgb}{1,0,0}
\definecolor{green}{rgb}{0,1,0}
\definecolor{blue}{rgb}{0,0,1}

\begin{document}
%%%%% title : short title may not be used but TITLE is required.
% \title{TITLE}
% \title[short title]{TITLE}
\title{Human mobility patterns at the smallest scales}

%%%%% author(s) :
% single author:
% \author[name in running head]{AUTHOR\corrauth}
% [name in running head] is NOT OPTIONAL, it is a MUST.
% Use \corrauth to indicate the corresponding author.
% Use \email to provide email address of author.
% \footnote and \thanks are not used in the heading section.
% Another acknowlegments/support of grants, state in Acknowledgments section
% \section*{Acknowledgments}
\author[Lind et.~al.]{Pedro G.~Lind\affil{1}\comma\corrauth,
                      Adriano Moreira\affil{2}}
\address{\affilnum{1}\ 
         ForWind - Center for Wind Energy Research, Institute of Physics,
         Carl-von-Ossietzky University of Oldenburg, DE-26111 Oldenburg, 
         Germany. \\
         \affilnum{2}\
         Centro Algoritmi, Escola de Engenharia, Universidade do Minho,
         Campus de Azur\'em, 4800-058 Guimar\~aes, Portugal}
\email{{\tt pedro.g.lind@forwind.de} (Lind),
       {\tt adriano@dsi.uminho.pt} (Moreira)}
% multiple authors:
% Note the use of \affil and \affilnum to link names and addresses.
% The author for correspondence is marked by \corrauth.
% use \emails to provide email addresses of authors
% e.g. below example has 3 authors, first author is also the corresponding
%      author, author 1 and 3 having the same address.
% \author[Zhang Z R et.~al.]{Zhengru Zhang\affil{1}\comma\corrauth,
%       Author Chan\affil{2}, and Author Zhao\affil{1}}
% \address{\affilnum{1}\ School of Mathematical Sciences,
%          Beijing Normal University,
%          Beijing 100875, P.R. China. \\
%           \affilnum{2}\ Department of Mathematics,
%           Hong Kong Baptist University, Hong Kong SAR}
% \emails{{\tt zhang@email} (Z.~Zhang), {\tt chan@email} (A.~Chan),
%          {\tt zhao@email} (A.~Zhao)}
% \footnote and \thanks are not used in the heading section.
% Another acknowlegments/support of grants, state in Acknowledgments section
% \section*{Acknowledgments}

%%%%% Begin Abstract %%%%%%%%%%%
\begin{abstract}
We present a study on human mobility at small spatial scales.
Differently from large scale mobility, recently studied through 
dollar-bill tracking and mobile phone data sets within one big 
country or continent, we report Brownian features of human 
mobility at smaller scales.
In particular, the scaling exponents found at the smallest 
scales is typically close to one-half, differently from the
larger values for the exponent characterizing mobility at
larger scales.
We carefully analyze $12$-month data of the Eduroam
database within the Portuguese university of Minho.
A full procedure is introduced with the aim of properly 
characterizing the human mobility within the network of access 
points composing the wireless system of the university.
In particular, measures of flux are introduced for estimating a
distance between access points. This distance is typically 
non-euclidean, since the spatial constraints at such small scales 
distort the continuum space on which human mobility occurs.
%Based in our findings, we discuss how to adapt our numerical
%procedure for deriving optimal wireless network topologies 
%avoiding local saturation at crowded access points.
Since two different exponents are found depending on the
scale human motion takes place, we raise the question at 
which scale the transition from Brownian to non-Brownian
motion takes place.
In this context, we discuss how the numerical approach can be 
extended to larger scales, using the full Eduroam in Europe and
in Asia, for uncovering the transition between both dynamical
regimes.
\end{abstract}
%%%%% end %%%%%%%%%%%

%%%%% AMS/PACs/Keywords %%%%%%%%%%%
\pac{%
%      89.75.Kd, %Pattern formation in complex systems, 
%      01.75.+m, %Science and society, 
%      89.90.+n} %Interdisciplinary physics, see section new topics in, 
       89.75.Da, %Systems obeying scaling laws
       89.75.Fb, %Structures and organization in complex systems
       89.75.Hc} %Networks and genealogical trees 

%02.50.Ga,  %Markov processes
%      02.50.Ey,  %Stochastic processes
%      92.70.Gt}   %Climate dynamics
%            92.60.Sz \sep   % Air pollution

%\ams{52B10, 65D18, 68U05, 68U07}

%%%% maketitle %%%%%
\maketitle

%%%%%%%%%%%%%%%%%%%%%%%%%%%%%%%%%%%%%%%%%%%%%%%%%%%%%%%%%%%%%%%%%%%%%%
%%%%%%%%%%%%%%%%%%%%% TEXT %%%%%%%%%%%%%%%%%%%%%%%%%%%%%%%%%%%%%%%%%%%
%%%%%%%%%%%%%%%%%%%%%%%%%%%%%%%%%%%%%%%%%%%%%%%%%%%%%%%%%%%%%%%%%%%%%%

%%%%%%%%
\section{Motivation and Scope}

Understanding human motion from small scales, such as buildings and
streets up to larger ones comprising cities, countries and continents,
has been proven to be important for a variety of application areas such 
as spread of diseases, 
opinion dynamics\cite{guo2010,toral2007},
and any other phenomena occuring on social networks\cite{ke2008},
as well as in optimization of telecommunication networks, urban 
planning or tourism management.
With these aims, several groups have been modeling human motion 
during the past few 
years\cite{gonzaleznature,brockmannnature,ahas2006,ahas2007}.

At middle and larger scales highly significant contributions 
for understanding human mobility have been made.
Brockmann and his team\cite{brockmannnature}, for example, 
have shown that human traveling distances within U.S.A. decay as a
power-law. On the other hand, Gonzalez and co-workers\cite{gonzaleznature} 
have shown that there is a high degree of temporal and spatial regularities
with a single probability distribution for returning to previous locations.
However a study occurring at lower scales, namely within one single building
or a small set of buildings still lacks to be addressed.
One of the reasons for this lays in the nature of available data.
For instance, Brockmann and his team\cite{brockmannnature} used 
data obtained from an online bill tracker system where registered users 
report the observation of marked US dollars bills around the United States, 
and a data set representing the position of “travel bugs” in GeoCaching 
systems. Though reasonable, these data sets however do not represent human
motion directly.
Gonz\'alez and co-workers\cite{gonzaleznature} proposed a way for tracking
human motion, using a data set containing positioning records of around 
$10^5$ users of a cellular network collected over a period of six months. 
Since mobile phones are personal devices, the trajectory of a mobile phone 
is highly correlated to that of his owner, turning to be a much better
proxy to observe the trajectories of humans than bills or other non-personal 
items. 
%%%%%%%%%%%%%%%%%%%%%%%%%%%%%%%%%%%%%%%%%%%%%%%%%%%%%%%%%%%%%%%%%%%%%%%
\begin{figure}[t]
\centering
\includegraphics[width=0.65\textwidth]{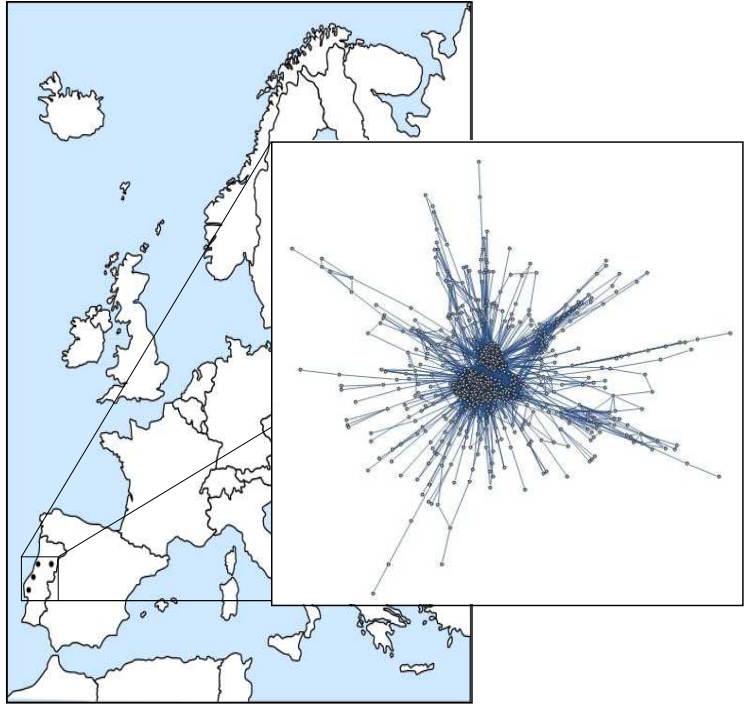}
\caption{\protect (Color online)
A snapshot of the Eduroam network comprehending
the data of University of Minho with its workers and students spreaded 
throughout other Eduroam networks in Portugal.
Time period spans from January till December 2011.}
\label{fig1}
\end{figure}
%%%%%%%%%%%%%%%%%%%%%%%%%%%%%%%%%%%%%%%%%%%%%%%%%%%%%%%%%%%%%%%%%%%%%%%

However, two drawbacks should be stressed in the data sets used so far.
First, in both cases the data records correspond to position data collected 
only when a person initiates/receives a phone call or a SMS message, or when
he or she declares online the dollar bill.
Second, there is the problem of spatial resolution. 
For example, for cell ID, the data 
cannot be used to validate the proposed models at short scales, since the 
coverage area of mobile network cells extends to several kilometers in 
rural areas.

In this paper we analyze a large data set of Eduroam networks at Portuguese 
Universities (see Fig.~\ref{fig1}), with the aim of addressing the problem 
of human motion at the smallest scales, i.e.~within one or a few buildings.
Our study follows from previous work\cite{adriano1,adriano2,adriano3}.
The data set comprises one year of collected information with a sample
frequency of one second, a time step that enables to assume the data as
continuous monitoring of human mobility, and with a higher spatial
resolution
than mobile phone calls data sets which enables the motion of human 
dynamics at its smallest scale. We will show that the statistical features 
of human mobility at the smallest scales is much closer to the random-walk 
model than the results obtained for studies at larger scales.
Still, the corresponding exponents we find in this study indicate a 
super-diffusive regime
which we address in the end for discussing possible ways of using such
statistical information to optimize Eduroam networks when establishing the
location of the access points.

The complex network framework has been applied frequently to address 
social and environmental problems\cite{lind05,gonzalez06,gonzalez06b,lind07,lind07b}.
For recent reviews see \cite{boccaletti06,dorogovtsev07}.
While using the same framework, we introduce here a novel procedure
for extracting the network on which persons move at small scales 
(university buildings), directly from empirical data.

We start in Sec.~\ref{sec:data} describing the data sets and in 
Sec.~\ref{sec:tools} we introduce tools to properly extract the network
of access points. As we will see, the procedure for extracting such network
is not trivial since the weight of each connection between two access points 
is not a straightforward euclidean distance due to the physical constraints
(stairs, walls, etc) that condition the movement of persons.
In Sec.~\ref{sec:mobility} a detailed description of our finding is presented,
enhancing the near-diffusive regime found for human mobility at small
scales. Further discussions on how to apply our findings for wireless network
optimization as well as the main conclusions are given in 
Sec.~\ref{sec:conclusions}.

%%%%%%%%
\section{The Eduroam data sets}
\label{sec:data}

The data set comprehends a Eduroam RADIUS log file from the University of
Minho (Portugal) infrastructure, including the full year of 2011.
The file contains a total of $15,892,009$ data records. From those, 
$7,937,245$ refer to ``start-events'' and $7,954,764$ refer to 
``stop-events''. 
For the present analysis, only records representing stop-events have been 
used, as they include the disassociation time-stamp and session time from 
where the start events can be computed. 
A few amount of these records ($0.65\%$) have been filtered, since they are 
incomplete or do not represent unique Access Points or unique Stations. 
After filtering, $7,902,828$ records remain for processing and analysis.

Using all these records we will, in the next section, construct the
weighted network composed by a number of Access Points 
(APs), that varies in time.
We constructed the AP network considering all APs detected during
each full month of 2011. The minimum number of APs was observed in August,
namely $723$ APs. 

For extracting the AP network,
we convert the Eduroam database into an output file
with three single fields for each record, namely:
\begin{itemize}
\item The user $i$, or alternatively the device one user is using 
when connected to the net. 
We will address the associated properties of users with lower-case letters.
\item The Access Point (AP) $I$.
Other properties of APs will also be addressed 
with capital letters.
\item Time $t$ in units of the
time step $\Delta t=1$ second, which is typically the size of
the smallest time lag between successive records.
\end{itemize}

%%%%%%%%%
\section{The underlying network and its evolution}
\label{sec:tools}

The global network we aim to construct will be extracted from two 
complementary networks included in our database.

One part comprehends the network of APs, a network composed solely by 
the APs, connected among them through weighted links with a weight 
$W_{IJ}$ inversely proportional to their topological distance $D_{IJ}$, 
that will be properly defined below.

The other part is the network of users, a network composed solely by the 
links between each connected user and the corresponding AP. 
The collection of these connections results in a completely disconnected set 
of APs each one with a number of users exclusively linked to them.

Notice that, the time one user is connected to one AP is quite larger
than the unit time-step (1 second). Therefore, by taking snapshots of the 
entire network at each increment $\Delta t=1$ second, the amount of data to 
analyze increases enormously, compared with the amount of data collected 
originally.
To overcome this shortcoming, the network of users is defined only by the 
linked-list mapping, at each time step, each connected user $i$ to its 
corresponding AP $I$.

\subsection{Access Point fitness and session times}

\begin{figure}[t]
\centering
\includegraphics[width=0.85\textwidth]{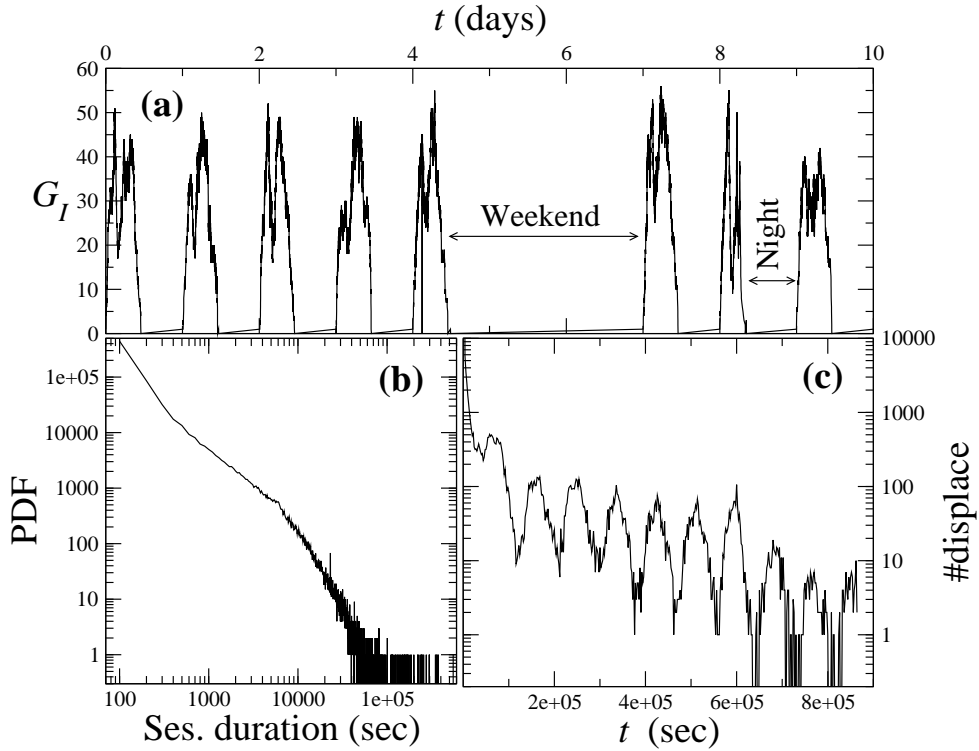}
\caption{\protect 
   {\bf (a)} Total number $G_I(t)$ of connected users as a function of 
             time,
   {\bf (b)} the distribution of the session time and
   {\bf (c)} the distribution of the
             duration of displacements indicating the existence of a
             daily period.}
\label{fig2}
\end{figure}

Counting the total number of connected users at one given time step
$t$ yields the fitness $G_I(t)$ of a particular AP $I$. 
See Fig.~\ref{fig2}a for a typical example on how $G_I(t)$ varies 
over a few days.

At each time-step $t$, $G_I(t)$ at AP $I$ equals the total number
of associations occurred at $I$ minus the total number of disassociations, 
occurred within the full time span from the beginning of the observation 
period up to time $t$.
Through time, the fitness illustrated in Fig.~\ref{fig2}a shows a 
daily quasi-periodic pattern, which indicates a maximum time interval 
of one day for the sessions of connected users.
Therefore, to avoid isolated associations we assume that each one cannot
last more that $24$ hours. Similar assumption can be taken for disassociations. 
Notice also that the initial value $G_I(0)$ for all $I$ is considered
to be zero, though it may correspond to same base value of connected
devices that varies from AP to AP.

%%%%%%%%%%%%%%%%%%%%%%%%%%%%%%%%%%%%%%%%%%%%%%%%%%%%%%%%%%%%%%%%%%%%%%%
%%%%%%%%%%%%%%%%%%%%%%%%%%%%%%%%%%%%%%%%%%%%%%%%%%%%%%%%%%%%%%%%%%%%%%%

From Fig.~\ref{fig2}b one can conclude that the large majority 
of sessions is shorter than one day, with less than $0.02\%$ of sessions
longer than $24$ hours.
Therefore, filtering sessions longer than 24 hours has no
significant impact on the following analysis and results.
Also from Fig.~\ref{fig2}c, showing the duration of displacements,
one observes a daily pattern. Here, 
an approximate fit of the distribution of session  
times is $m(x) = a \exp{(-x^b)}$ with $a\sim 4\times
10^5$ and $b\sim 0.23$ (not shown).

%These two paragraphs are not clear. Both seem to stress the same 
%idea: sessions are shorter than one day. Also, the reference to 
%Fig. 2c is confusing since it displays information about the temporal 
%duration of the displacements, not the sessions. I know they are 
%somehow correlated, but that is not apparent now.

Displacements between APs, defined as the disassociation from 
one AP $I$ and subsequent association with AP $J$, exhibit a periodic 
behavior as illustrated by the distribution of the corresponding 
time spans (see Fig.~\ref{fig2}c). However, the majority of the 
displacements have a short duration, representing fast displacements 
among nearby APs.
%%%%%%%%%%%%%%%%%%%%%%%%%%%%%%%%%%%%%%%%%%%%%%%%%%%%%%%%%%%%%%%%%%%%%%%
    \begin{figure}[t]
        \centering
        \includegraphics[width=0.48\textwidth]{./fig03ab_eduroam.eps}
        \includegraphics[width=0.48\textwidth]{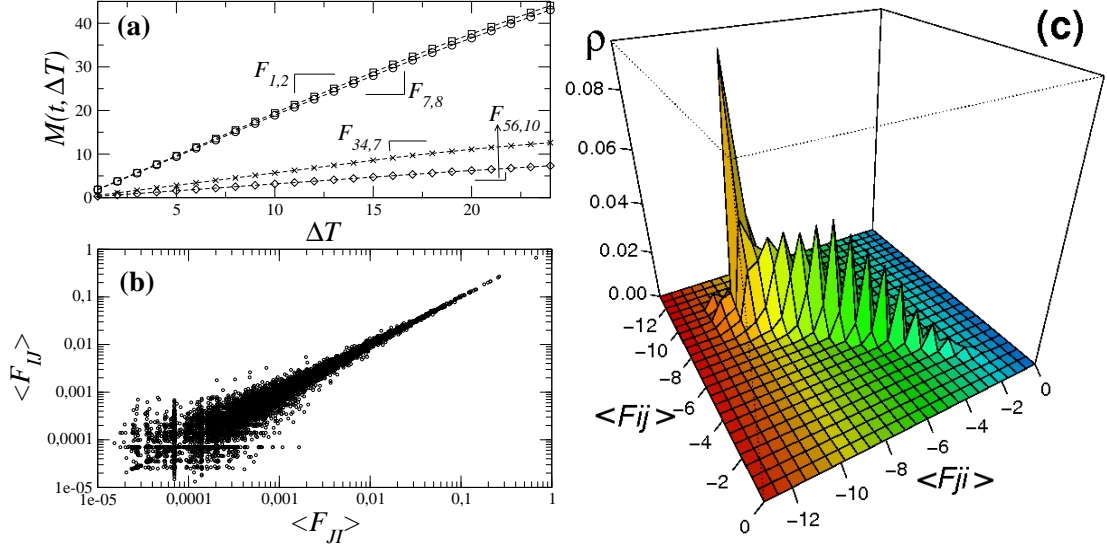}
        \caption{\protect 
          {\bf (a)} 
          Number $M$ of AP switches as a function of the time 
          window $\Delta T$, defining the flux 
          $F_{IJ} =M/\Delta T \sim dM/d(\Delta T)$ 
          (see Eq.~(\ref{instflux})) which 
          is approximately constant for $\Delta T$.
          Shown is the slope for four different pairs $IJ$
          of APs taken at the first day of 2011.
          {\bf (b-c)} Plot of average fluxes between each pair of APs
          in both directions, showing a symmetry relation 
          $\langle F_{IJ}\rangle \simeq \langle F_{JI}\rangle$ 
          which supports the definition for
          the (symmetric) weights $W_{IJ}$ in the AP network 
          (see text).}
        \label{fig3}
    \end{figure}
%%%%%%%%%%%%%%%%%%%%%%%%%%%%%%%%%%%%%%%%%%%%%%%%%%%%%%%%%%%%%%%%%%%%%%%
%%%%%%%%%%%%%%%%%%%%%%%%%%%%%%%%%%%%%%%%%%%%%%%%%%%%%%%%%%%%%%%%%%%%%%
    \begin{figure}[htb]
        \centering
        \includegraphics[width=0.48\textwidth]{./fig04a_eduroam.eps}
        \includegraphics[width=0.45\textwidth]{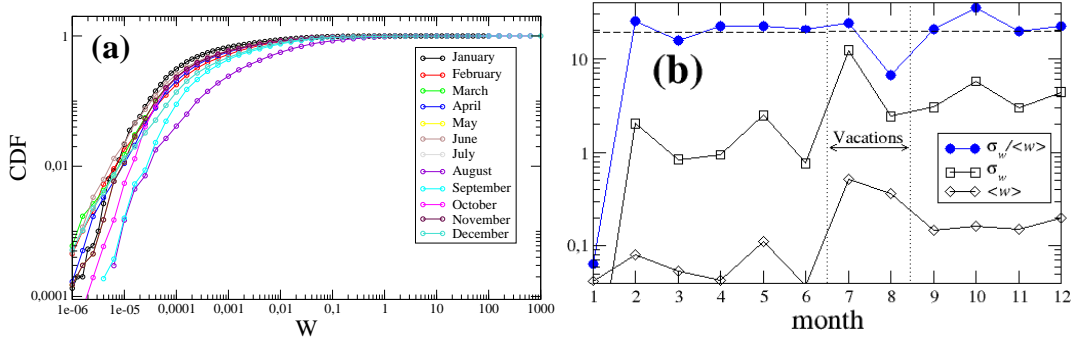}
        \caption{\protect 
          {\bf (a)} Cumulative density distribution $P(W)$ for the
          weights computed in each month of 2011 (see text) and
          {\bf (b)} Weight average $\langle W\rangle$ and standard 
          deviation $\sigma_W$ as a function of 
          month, using January as the reference month.}
        \label{fig4}
    \end{figure}
%%%%%%%%%%%%%%%%%%%%%%%%%%%%%%%%%%%%%%%%%%%%%%%%%%%%%%%%%%%%%%%%%%%%%%%
%%%%%%%%%%%%%%%%%%%%%%%%%%%%%%%%%%%%%%%%%%%%%%%%%%%%%%%%%%%%%%%%%%%%%%%
\begin{figure*}[htb]
   \centering
   \includegraphics[width=0.85\textwidth]{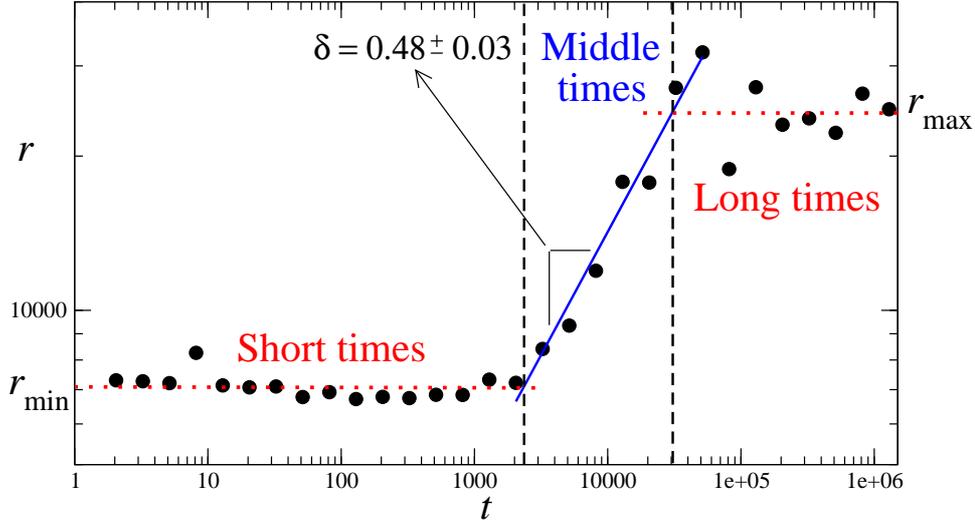}
   \caption{\protect 
   Average displacement $r$ (bullets) of human trajectories within Portuguese
   Universities as a function of time (in seconds). While for
   short times people tend to remain at the same place (connected to
   the same AP), with $r_{min}$, beyond around $30$ minutes there is
   a marked scaling of the displacement with time, $r\sim t^\delta$
   with $\delta \simeq 0.48\pm 0.03$ beyond the normal Brownian diffusion
   regime. Beyond the eight working hours the displacement stabilizes 
   at a maximum value $r_{max}$.} 
   %At middle times the average displacement here computed
   %resembles the radius of gyration $r_g$ (circles) taken in 
   %previous studies\cite{gonzaleznature} (see text).}
\label{fig5}
\end{figure*}
%%%%%%%%%%%%%%%%%%%%%%%%%%%%%%%%%%%%%%%%%%%%%%%%%%%%%%%%%%%%%%%%%%%%%%%

\subsection{Fluxes between Access Points and their topological 
distance}

The network of APs is a weighted network and its adjacency matrix 
is defined by the elements $W_{IJ}=1/D_{IJ}$, where $D_{IJ}=D_{JI}$ is 
the distance between APs $I$ and $J$.
Typically we do not have the location of the APs and we do not know
exactly the full number of constraints that condition the 
displacement of a person from the coverage area of one AP to that of 
the next one.
Therefore we introduce here a definition
for the topological distance between APs in such networks.

To that end we need the average fitness $\langle G_I(t) \rangle$ at 
each AP $I$ and the flux of users that go from AP $I$ to AP $J$, without 
passing in any intermediate AP.
We notice that seasonal fluctuations have been observed while using
the network, with the largest variations over the summer (vacations) period
(see Fig.~\ref{fig4} below). Therefore, the average fitness at each AP 
has been calculated for a time window of $1$ month and repeated for every 
month during the entire year. 
A similar approach has been used for the calculation of the average flux, 
which yields one AP network for each month.

For computing the flux from $I$ to $J$, one reads the output file counting
the number $M$ of users that associate to AP $J$ after disassociating 
from AP $I$, within a time window $\Delta T \lesssim 1$ day.
The instantaneous flux $F_{IJ}$ between $I$ and $J$ is then defined as 
\begin{equation}
F_{IJ}(t)=\frac{dM(t,\Delta T)}{d(\Delta T)} .
\label{instflux} 
\end{equation}
From the instantaneous flux the average flux is easily computed as
\begin{equation}
\langle F_{IJ}\rangle = \frac{1}{N_{days}}\int_0^{N_{days}}F_{IJ}(t)dt .
\label{aveflux}
\end{equation} 
As shown in Fig.~\ref{fig3}a, for four different pairs $IJ$ of APs, 
the average flux is approximately
constant within time windows up to one day. Therefore we consider
the average flux within $\Delta T=4$ hours as an estimate of
the flux between each pair of APs to reduce the computation time. 
%In the latest results I computed <FIJ> for the 24 hours to avoid the 
%cases where the use of the network is more during the morning, more 
%during the afternoon, etc.
%Anyway, in most of the cases, computing for dT = 4 hours is enough, 
%so we can leave this as it is.
%I NEED TO CHECK THIS
For a fixed $\Delta T$, the average flux 
depends on $t$: computing Eq.~(\ref{aveflux}) in different days
or weeks yields different results.
Still, preliminary results (not shown) indicate that for several
pairs of APs the variation of $\langle F_{IJ}\rangle$ is not 
very large. Henceforth, we make two assumptions:
(i) the fluxes do not vary significantly within one month
and (ii) they are well represented through the computation
of Eq.~(\ref{aveflux}) for $N_{days}=7$ days.

Moreover, as shown in Fig.~\ref{fig3}b the fluxes are typically symmetric,
i.e.~$\langle F_{IJ}\rangle=\langle F_{JI}\rangle$.
From these quantities $\langle F_{IJ}\rangle$ and $\langle G_{I}(t)\rangle$ 
one finally defines the weight of the connection between AP $I$ and
$J$ as follows:
\begin{equation}
\frac{1}{D_{IJ}} = \tfrac{1}{2} \left (
       \frac{\langle  F_{IJ}(t)\rangle}{\langle  G_{I}(t)\rangle}+
       \frac{\langle  F_{JI}(t)\rangle}{\langle  G_{J}(t)\rangle}
     \right ) \equiv W_{IJ} = W_{JI} .
\label{eqFlux}
\end{equation}
Here $\langle G_I\rangle$ is average over one month.

Figure \ref{fig4}a shows the weight distribution (cumulative density
functions) for each month in the data set.
The weight average $\langle W \rangle$ -- over all AP connections $IJ$ -- 
varies from month to month as well as the corresponding deviations 
\begin{equation}
\sigma_W(t) = \left ( \frac{1}{N-1}\sum_{connections} 
\left ( W_{IJ}(t)-\langle W\rangle (0)\right )^2\right )^{1/2}
\end{equation}
from a reference month, chosen to be January.
Figure \ref{fig4}b shows the evolution through one year of both the
average (diamonds) and the deviation (squares) from the average 
$\langle W\rangle$. While some
significant variations are observed during the summer vacations,
the normalized deviations $\sigma_W/\langle W\rangle$ are constant.
This result also highlight the temporal evolution on the usage of the 
network, with more and more users accessing the network and also 
more APs being deployed to improve the network capacity.

%%%%%%%%%
\section{Mobility patterns at small scales}
\label{sec:mobility}

To study the mobility within the universities we do as follows.
First, we make use of the 
inverse weight of the connections between adjacent APs
to define a topological distance. Since the 
weight measures the normalized flux between two adjacent APs, say 
$I$ and $J$, its inverse value can be taken as a good estimate of 
the topological distance $D_{IJ}$ between those APs, which of course 
defines a symmetric matrix ($D_{IJ}=D_{JI}$).
Then, having defined the matrix of topological distances between APs,
we symbolize the position of one person at time $t$ throughout its 
trajectory $i$ as 
$r_i(t)$ with reference to an initial position, labeled AP $I_0$.
In other words, if person $i$ at time $t$ is connected to
AP $I(t)$, its position is $D_{I_0I(t)}$. % (see Eq.~(\ref{eqFlux})).}

Second, having the trajectory $i$ followed by one person defined by 
the successive values of APs, $I(t)$, and also the corresponding distances
$D_{I_0I(t)}$ to the initial AP $I_0$, we are able to compute the
average distance $r(t)$ to the initial position as a function of time.
Indeed, keeping track of the distance from the initial position, and
averaging over all trajectories in the data sets yields the
average distance from the initial position:
\begin{eqnarray}
r(t)&=&
\frac{1}{N_{traj}(n)}\sum_{k=1}^{N_{traj}(n)} D_{I_0I(t_0+n\Delta t)},
\label{eq:r}
\end{eqnarray}
with $i$ labeling one of the $N_{traj}(n)$ trajectories observed at each time
$t$ and $t=t_0+n\Delta t$ where $t_0$ marks the initial time when the 
trajectory is started and we chose $\Delta t=1$ second.

Figure \ref{fig5} shows the displacement $r(t)$ as a function of
time $t$ averaged over all $N_{traj}$ observed in year 2011. 
While for
short and long times the displacement tends to remain approximately
constant at $r_{min}$ and $r_{max}$ respectively, for middle times
there is a clear scaling $r\sim t^\delta$ with $\delta= 0.48\pm 0.03$.
Short times are shorter than $30$ minutes: sessions are too short
for observing any significant motion from one AP to the next
one. Long times are typically longer than $\sim 8$ hours, which
comprehends the typicall working day of researchers and students
in the Eduroam network.

%%%%%%%%%%%%%%%%%%%%%%%%%%%%%%%%%%%%%%%%%%%%%%%%%%%%%%%%%%%%%%%%%%%%%%%
\begin{figure}[htb]
   \centering
   \includegraphics[width=0.85\textwidth]{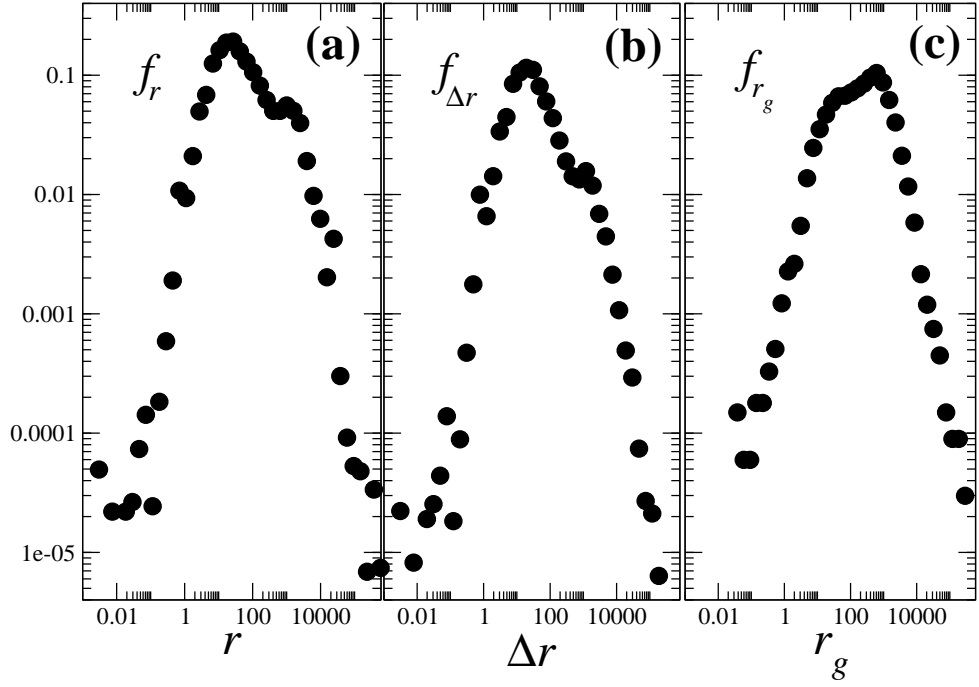}
   \caption{\protect 
      Distribution of the displacements $r$, its increments 
      $\Delta r$ and the radius of gyration $r_g$ (see text). 
      Typically one finds typical values for each one of these 
      variables, differently from the power-law tails found 
      in human motion at larger scales\cite{gonzaleznature} (see 
      Eqs.~(\ref{eq:rgpdf})).}
\label{fig6}
\end{figure}
%%%%%%%%%%%%%%%%%%%%%%%%%%%%%%%%%%%%%%%%%%%%%%%%%%%%%%%%%%%%%%%%%%%%%%%

In the study by Gonzalez and co-workers\cite{gonzaleznature}, instead
of the average distance $r(t)$ a radius of gyration was used.
The radius of gyration $r_g(t)$ is defined as the linear size occupied
by each trajectory up to time $t$ and is computed as
\begin{equation}
r_g(t)=\left (    
       \frac{1}{N_{traj}(t)} \sum_{i=1}^{N_{traj}(t)} (r_i(t)-r_i^{(cm)}(t))^2
       \right )^{1/2}
\end{equation}
where $r_i{(cm)}(t)$ is the ``center of mass'' of trajectory $i$
at time $t$:
\begin{equation}
r_i^{(cm)}(t)=
       \frac{1}{N_{traj}(t)} \sum_{i=1}^{N_{traj}(t)} r_i(t) .
\end{equation}
The reason why we chose to use the average displacement instead of 
the radius of gyration lays in the fact that the former is less
sensitive to the transition from short times to middle times.
%In Fig.~\ref{fig5} we indicate with circles the radius of gyration.
For short times the radius of gyration is constant and significantly
larger than the average distance. When the trajectory starts to leave
the original AP the center of mass $r_i^{cm}(t)$ of the trajectory
increases which leads to a decrease of the radius of gyration.
Finally, when the center of mass stabilizes one observes the similar
behavior between our average distance $r(t)$ and the radius of gyration
introduced previously by Gonzalez et al\cite{gonzaleznature}.
For long times the radius of gyration stationarizes also approximately 
at the same value as the average distance.

Having obtained the value of $\delta$ for the average distance
%(or alternatively for the radius of gyration) 
one can now proceed to
access dynamical features of these human trajectories at small scales.

In general, the exponent $\delta$ characterizes how people ``diffuse'' 
in space. For $\delta=1/2$ one has the normal Brownian diffusion, which
can be describe by a person moving continuously in a random direction
in small jumps. If this walker moves slower, with longer waiting times
then $\delta<1/2$. On the contrary, if the walker is able to perform
large jumps while randomly walking, then $\delta>1/2$.
Typically, $\delta=1/2$ characterizes the so-called normal diffusion,
while $\delta <1/2$ and $\delta >1/2$ indicate anomalous regimes,
sub-diffusive and super-diffusive respectively. For a short review
in this topic see Ref.~\cite{shlesinger99}.

Based in this knowledge one observes that the exponent $\delta=0.48\pm 0.03$
is the same as the one for Brownian motion, within numerical error,
differently from the exponent found in human mobility studies at larger 
scales\cite{gonzaleznature}.
Indeed in the study by Gonzalez the authors
found for mobility within U.S.A.~an exponent $\beta_r=1.65$ 
for the distribution of radius of gyration
\begin{equation}
f(r_g)=\left ( r_g+r_g^0\right )^{-\beta_r} \exp{\left ( -r_g/\kappa\right )} .
\label{eq:rgpdf}
\end{equation}
Since it is known from scaling analysis that $r_g\sim t ^{3/(2+\beta_r)}$, one
finds $\delta_{large}=3/(2+\beta_r)=0.82$ for the large spatial scales covered
in \cite{gonzaleznature}. This value is quite above the value found for
the mobility within universities.

Moreover, the functional form of the $r_g$ distributions given
in Eq.~(\ref{eq:rgpdf}) is quite different from the one observed in
our study. In particular, the power-law tail is for our small spatial
scales not observed.
Figure \ref{fig6} shows the distribution $f(r)$ together with the
distribution $f(r_g)$ for the radius of gyration and also 
the distribution $f(\Delta r)$ of the displacement increments
$\Delta r(t) = r(t+\Delta t)-r(t)$.
In all cases the distribution shows no power-law tendency. In particular
it is characterized by a typical value for $r$, $r_g$ and $\Delta r$,
resembling therefore much of the Brownian behavior occurring in 
normal diffusion.

%%%%%%%%%%%%%%%%%%%%%%%%%%%%%%%%%%%%%%%
\section{Discussion and conclusions}
\label{sec:conclusions}

The central finding in this paper is the $\delta$-value found for human
motion at the smallest scale (streets and buildings). This value is close 
to the one for normal diffusion regime, much lower than the value 
found in \cite{gonzaleznature} for middle and large scales, cities and countries
respectively. What does this mean?

At large scales such as countries it is known that people make large 
``jumps'' with a significant frequency. This is reflected in a large $\delta$
value, considerably above one. As one decreases to smaller scales this
large jumps turn to occur more rarely. At the most extreme case, one can
think in a set of people moving on a continuous surface not larger than a 
few hundred meters wide. In this situation the average displacement exponent 
should be $1/2$. Within a building however there are physical constraints, 
such as stairs and walls, that do not promote a normal diffusion of persons. 
Some deviation from the Brownian exponent would be expected.
Still, our procedure for extracting the topological distance between
pairs of APs seems to properly account for such spatial constraints in
human motion at small scales.

Of course that in our case the position are determined by the locations of
the AP. If one changes their location the AP network would be different and,
for the same human dynamics, the exponent would be also different.
A set of locations corresponding to an exponent $\delta=1/2$ is conjectured
to be the optimal set of locations, i.e.~the one that better suits the 
flux of persons throughout the building.

Therefore, with the approach we introduce in this paper, one is able to
construct a weighted network of APs and evaluate how well it suits the
mobility of persons where the APs are placed. 

As recently proposed\cite{raischel2014},
since the Eduroam data base covers the whole of Europe, a possible 
next step would be to consider the joint data base of several 
European universities. 
By properly matching the mask IDs and similar 
quantities at different universities, one would be able to keep
track of individuals throughout Europe and apply the framework
described above to data sets comprising larger and larger
areas. Since the mobility of faculties and students is becoming
higher due to European programs such as Erasmus, one
should expect sufficient statistics 
in the Eduroam databases at several spatial scales.
Consequently, one should be able to
observe the transition from Brownian to non-Brownian motion,
postulated in this paper.

In particular, for the entire region of Asia, eduroam is deployed with
top-level RADIUS servers beeing operated by AARNet (Australia) and
by the University of Hong Kong. Since this is the broadest region
having Eduroam operating, covering the full area from New Zealand 
to Japan and from South Korea to India, it is of particular interest
for assessing the full set of scales where human motion takes place.

Two further points that would be interesting to explore in these data sets
are, first, the time evolution of the network of AP, here taken as a 
static network, and, second, the multifractality of human dynamics at 
small scales, i.e.~to
study the dependence of exponent $\delta$ on the order of the moments
of the displacement $r$: $\langle r^q\rangle \sim t^{\delta(q)}$.
These and other points will be addressed elsewhere.

%%%%%%%%%%%%%%%%%%%%%%%%%
\section*{Acknowledgments}

The authors thank
partial support by {\it Funda\c{c}\~ao para a Ci\^encia e a Tecnologia}
under the R\&D project PTDC/EIA-EIA/113933/2009.
PGL also thanks support from PEst-OE/FIS/UI0618/2011.

%%%%%%%%%%%%%%%%%%%%%%%%%%%%%%%%%%%%%%%%%%%%%%%%%%%%%%%%%%%%%%%%%%%%%%%%%%%
%%%%%%%%%%%%%%REFERENCIAS %%%%%%%%%%%%%%%%%%%%%%%%%%%%%%%%%%%%%%%%%%%%%%%%%
%%%%%%%%%%%%%%%%%%%%%%%%%%%%%%%%%%%%%%%%%%%%%%%%%%%%%%%%%%%%%%%%%%%%%%%%%%%

\end{document}